\documentclass[prd,twocolumn,preprintnumbers]{revtex4}
 \usepackage{amsmath,amssymb,bm}
\renewcommand\d{\partial}
\newcommand\x{\mathbf x}
\newcommand\y{\mathbf y}
\newcommand\cRL{c^{\phantom{I}}_{\mathrm{RL}}}

\begin{document}
\preprint{EFI-14-24}
\title{Thermal Transport in a Noncommutative Hydrodynamics}
\author{Michael Geracie and Dam Thanh Son}
\affiliation{Kadanoff Center for Theoretical Physics, University of
Chicago, Illinois 60637, Chicago, USA}
\begin{abstract}
We find the hydrodynamic equations of a system of particles
constrained to be in the lowest Landau level.  We interpret the
hydrodynamic theory as a Hamiltonian system with the Poisson brackets
between the hydrodynamic variables determined from the
noncommutativity of space.  We argue that the most general
hydrodynamic theory can be obtained from this Hamiltonian system by
allowing the Righi-Leduc coefficient to be an arbitrary function of
thermodynamic variables.  We compute the Righi-Leduc coefficient at
high temperatures and show that it satisfies the requirements of
particle-hole symmetry, which we outline.

\begin{flushright}
  \emph{Contribution for the JETP special issue in honor of
    V.~A.~Rubakov's 60th birthday}
\end{flushright}
\end{abstract}
\maketitle

\section{Introduction.}

Interacting electrons in very high magnetic
fields show extremely rich behaviors, the most well-known of which is
the fractional quantum Hall (FQH)
effect~\cite{Tsui:1982yy,Laughlin:1983}.  In the most interesting
limit, all the physics occurs in the lowest Landau level (LLL) and
originates from the interactions.

In this paper, we study the finite-temperature dynamics of electrons
in a magnetic field so high that all particles are constrained to be
on the LLL.  This problem is the finite-temperature counterpart of the
FQH problem.  While many quantum phenomena are smeared out by the
temperature, the hydrodynamic theory, which takes hold at distances 
and time scales much larger than the mean free path/time, is expected 
to be universal.  We
assume the system is clean, without impurities, and the only
relaxation mechanism is the interactions between particles.
This regime is particularly relevant for the proposed realizations of 
the 
FQH regime in cold atomic 
gases~\cite{Cooper:2008,Cooper:2013,Yao:2013}
length/time.  The main outcome of our investigation is the set of
hydrodynamic equations [Eqs.~(\ref{hydro-eqs}), (\ref{j}), and
  (\ref{ei})] which describes the long-wavelength dynamics of the
system, the identification of the kinetic coefficients, and the
computation of the thermal Hall coefficient in the high-temperature
regime [Eq.~(\ref{cRL-hiT})].

Previous studies of transport in high magnetic field include
Refs.~\cite{Zyryanov:1964,Obraztsov:1965,SmrckaStreda:1977,CHR:1997}.
In particular, in Ref.~\cite{CHR:1997} a general approach based on
conservation laws is developed for the hydrodynamics of a system in a
quantizing magnetic field.  This is the approach that we will follow
in this paper.  We concentrate here, however, on the LLL limit (the
zero mass limit), which should be a regular limit when the particle
carries a magnetic moment corresponding to the gyromagnetic factor 
$g=2$.  This
allows us to consider the response of the system to variations of the
magnetic field, as well as to discuss the particle-hole symmetry of
the hydrodynamic equations.

An important concept in our discussion is that a particle in the
lowest Landau level effectively lives on a noncommutative
space~\cite{Girvin:1986zz}, with its two coordinates $x^1$, $x^2$
satisfying the commutation law $[x^1,x^2]=-i\ell_B^2$.  This idea has
attracted some attention in the context of the quantum Hall effect; it
has been speculated that the appropriate field theory of the quantum
Hall effect should be a noncommutative field theory (see, e.g.
Refs.~\cite{Susskind:2001fb,Fradkin:2002qw}.  For an introduction to
noncommutative field theory, see Ref.~\cite{Rubakov_noncomm}.)  We use
this noncommutativity to argue for a particular Poisson bracket
algebra between hydrodynamic variables, and proceed to derive the
hydrodynamic equations from the Poisson brackets with the Hamiltonian.
This approach is inspired by the Hamiltonian formulation of classical
hydrodynamics~\cite{Landau:1941}.  The Hamiltonian equations that
follow from the formalism form a self-consistent hydrodynamic theory,
but we will argue that they need a slight modification to become the
most general set of equations consistent with conservation laws and
the second law of thermodynamics.  This modification is related to the
Righi-Leduc (thermal Hall) effect~\cite{Kane:1997fda,Stone:2012ud}.

\section{Thermodynamics and conservation laws}

Let us recall the basic thermodynamic
functions of a system in an external magnetic field~\cite{CHR:1997}.
The grand potential is an extensive thermodynamic variable which
depends on the temperature, chemical potential, and magnetic field:
$\Omega=-VP(T,\mu,B)$.  The partial derivatives of $P$ are the entropy
density, particle number density and magnetization:
$dP=sdT+nd\mu+MdB$.  The hydrodynamic pressure is not $P$ but its
Legendre transform with respect to $B$: $p=P-MB$, and hence
$dp=sdT+nd\mu-BdM$.  The energy density is $\varepsilon=Ts+\mu n-P$.

We consider a system of nonrelativistic particles
of mass $m$ and gyromagnetic factor $g=2$ moving in a background
magnetic field $B$, and will be interested in the regime where all
higher Landau levels can be neglected.  We will study the response of
the system to arbitrary fluctuations of both electric and magnetic
fields, assuming that $B$ does not vanish at any place in space
and time so that the separation between the lowest and the higher
Landau levels is always maintained.  The LLL limit corresponds to
taking $m\to0$ and all the physics should be finite in this limit for
$g=2$.

The Hamiltonian for our system is
\begin{multline}\label{H}
  H = \int\!d\x \left[ \frac{|D_i\psi|^2}{2m} - \Bigl(A_0+ \frac 
B{2m}\Bigr)
      \psi^\dagger\psi\right]+ \\ + \mathrm{interactions},
\end{multline}
where $D_i=\d_i-iA_i$ (we use units where $\hbar=c=1$ and absorb
the electron charge $e$ into the gauge potential $A_\mu$).  We can
also think about our system as that of particles with zero magnetic
moment ($g=0$), subjected to an external field in which the scalar
potential is tuned to deviate from $B/2m$ by an amount which remains
finite when $m\to0$.  The conservation laws are the conventional ones,
with the replacement $A_0\to A_0+B/2m$,
\begin{align}
  & \frac{\d n}{\d t} + \d_i j^i = 0, \\
  & \frac{\d}{\d t}(m j^i) + \d_k \Pi^{ik} = n\Bigl(E_i + \frac{\d_i 
B}{2m}
    \Bigr) + \epsilon^{ik}j_k B,\\
  & \frac{\d\varepsilon}{\d t}+\d_i \varepsilon^i = j^i \Bigl(E_i 
       + \frac{\d_i B}{2m} \Bigr). 
\end{align}
We now extract the part divergent at $m\to0$ from the conserved 
currents
and the stress tensor in the following manner,
\begin{align}
  \tilde j^i &= j^i + \frac{\epsilon^{ij}}{2m}\d_j n, 
\label{jjtilda}\\
  \tilde \Pi^{ik} &= \Pi_{ik} + \frac12(\epsilon^{ij}\d_j\tilde j^k
    + \epsilon^{kj}\d_j\tilde j^i) - \frac{nB}{2m}\delta^{ik},\\
  \tilde\varepsilon &= \varepsilon -\frac{nB}{2m}\,,\quad
  \tilde\varepsilon^i = \varepsilon^i - \frac1{2m}(Bj^i 
    + \epsilon^{ij}E_j n).
\end{align}
For the number current~(\ref{jjtilda}), this procedure of extracting
the $1/m$ part was done in Ref.~\cite{Simon:1996}.  The conservation
laws are regular in the $m\to0$ limit in terms of the newly defined
quantities,
\begin{subequations}\label{hydro-eqs}
\begin{align}
  & \frac{\d n}{\d t} + \d_i \tilde j^i = 0, \\
  & \frac{\d}{\d t}(m\tilde j^i) + \d_k \tilde \Pi^{ik} =
     nE_i + \epsilon^{ik}\tilde j_k B,\label{momentum-cons}\\
  & \frac{\d\tilde\varepsilon}{\d t} + \d_i \tilde\varepsilon^i =
     \tilde j^i E_i . \label{energy-cons}
\end{align}
\end{subequations}
These equations can also be obtained within the Newton-Cartan
formalism~\cite{GSWW}.  Moreover, in the limit $m\to0$ the first term
in the left-hand side of Eq.~(\ref{momentum-cons}) can be dropped, and
it becomes a force-balance condition.  From now on we will drop the
tildes in the finite currents.  To close the equations we need to
express $j^i$, $\varepsilon^i$ and $\Pi^{ik}$ through the derivatives
of the local temperature and chemical potential.  To first order in
derivatives in the equations, we can limit ourselves to the leading
order contribution to the stress tensor: $\Pi^{ik}=p\delta^{ik}$.

\section{Hamiltonian model of a noncommutative fluid}

We start with a
simple model of particles moving in the lowest Landau level.  We
number the particles by the index $A=1\ldots N$, and the
spatial coordinates by $i$.  The coordinates of a particle do not commute
with each other, but commute with those of other particles,
\begin{equation}
  \{ x_A^i,\, x_B^j \} = \delta_{AB} \frac{\epsilon^{ij}}{B(\x_A)}\,.
\end{equation}
The particle number density,
\begin{equation}
  n(\x) = \sum_A\delta(\x-\x_A),
\end{equation}
then has the following Poisson bracket,
\begin{equation}
  \{ n(\x),\, n(\y)\} = 
  -\epsilon^{ij} \d_i\left(\frac n B\right)\d_j\delta(\x-\y).
\end{equation}

We now need to understand the Possion brackets involving the entropy
density.  Recall that in ideal hydrodynamics the entropy per particle
$s/n$ is conserved along fluid worldlines.  We can assume that each
particle $A$ carries an entropy $s_A$ for all time,
\begin{align}
  s(\x) = \sum_A s_A \delta ( \x - \x_A). \label{entropy}
\end{align}
as $s'$ is in the continuum picture. 
We find
\begin{align}
  \{s(\x),\, n(\y)\} &= -\epsilon^{ij} \d_i \left( \frac{s}{B}\right)
    \d_j \delta(\x-\y), \\
  \{s(\x),\, s(\y)\} &= -\epsilon^{ij}\d_i c\,
    \d_j \delta(\x-\y),
\end{align}
where
\begin{align}
  c = \sum_A \frac{s_A^2}{B(\x_A)} \delta ( \x - \x_A ).
\end{align}
In order to close the Poisson algebra, we should express $c$ in terms
of $s$ and $n$.  In the ``mean field'' approximation we may expect
$c=s^2/nB$.  We shall for now assume the most general $c$ compatible
with the Jacobi identity, which can be shown to be
\begin{equation}\label{Jacobi}
  c = \frac nB f\Bigl(\frac sn\Bigr).
\end{equation}

Now the hydrodynamic equations can be obtained by computing Poisson
brackets with the Hamiltonian
\begin{equation}
  H = \int\!d\x\, \bigl[\varepsilon(s(\x),n(\x),B(\x)) - A_0(\x) 
n(\x)\bigr].
\end{equation}
Note that both the total particle
number and the total entropy are Casimirs of the Poisson algebra, so
they are automatically conserved.  We find, for example, $\d_t n=-\d_i
j^i$ where the particle number current $j^i$ is
\begin{equation}\label{j}
  j^i = \frac{\epsilon^{ij}}B \left[ n(E_j-\d_j\mu)-s\d_j T\right] 
  + \epsilon^{ij}\d_j \alpha.
\end{equation}
where $\alpha$ cannot be determined from charge conservation
alone. This can be done using the force balance
equation~(\ref{momentum-cons}), into which we substitute
$\Pi_{ik}=p\delta_{ik}$,
\begin{equation}
  \d_i p = n E_i + \epsilon^{ik}j_k B,
\end{equation}
which, by using $dp=sdT+nd\mu-BdM$ completely determines $j^i$, and
the result corresponds to $\alpha=M$.  The first term on the right-
hand side
of Eq.~(\ref{j}) corresponds to the ``transport current,'' while the
second part is the ``magnetization current.''

Computing the Poisson bracket of $s$ with the Hamiltonian, we can find
the conservation law for the entropy,
\begin{equation}\label{si}
   \d_t s + \d_is^i = 0, \quad 
   s^i = \epsilon^{ij} \left[ \frac sB(E_j-\d_j\mu) -c \d_j T\right].
\end{equation}
For energy density, we can use $\d_t\varepsilon=T\d_t s+\mu\d_t n$ and
derive from Eq.~(\ref{energy-cons})
the energy current
\begin{multline}\label{eps-i}
  \varepsilon^i = \epsilon^{ij}\Bigl[ \frac{\varepsilon+p}B (E_j-
\d_j\mu) 
  - M\d_j\mu - \\ - \Bigl(\frac{\mu s}B + cT \Bigr)\d_j T + \d_j M_E\Bigr].
\end{multline}
We have also introduced the ``energy magnetization'' $M_E$ whose
contribution to the energy current is divergence-free.

\subsection{St\v reda formulas}

We note here in passing that the St\v
reda formula can be derived from our equation for the current.
Expanding the current in terms of derivatives of thermodynamic 
variables,
including the derivative of $B$,
\begin{equation}
  j^i = \epsilon^{ij} (\sigma_H E_j + \sigma_H^\mu \d_j\mu
        + \sigma_H^T \d_j T + \sigma_H^B \d_j B),
\end{equation} 
we can then read out, for example
\begin{equation}
\begin{split}
  \sigma_H &= \frac nB\,,\\
  \sigma^\mu_H &= -\frac nB + \left(
    \frac{\d M}{\d\mu}\right)_{TB}
  = -\frac nB + \left(
    \frac{\d n}{\d B}\right)_{\mu, T}
\end{split}
\end{equation}
where we have used a Maxwell's relation.  
The naive Einstein relation $\sigma^\mu_H=-\sigma_H$ does not hold due
the nonvanishing magnetization current in thermal equilibrium.  Note
that in a zero-temperature incompressible phase $n/B$ is constant and
$\sigma^\mu_H=0$, consistent with the expectation that small spatial
variations of $\mu$ should not have any physical effect in such a
phase.  In thermal equilibrium the chemical potential traces the
electric field, $\mu=A_0$, and so the only current flowing in the
system is the magnetization current, equal to
$j^i=\sigma_H^{\rm eq} \epsilon^{ij} E_j$ where
\begin{equation}
  \sigma_H^{\rm eq} = \sigma_H + \sigma^\mu_H 
  = \left( \frac{\d n}{\d B}\right)_{\mu, T} . 
\end{equation}
This is the St\v reda formula~\cite{Streda:1982}. 

The thermopower can be read out from our expression for the transport
current: it is equal to entropy per particle $s/n$, a known
result~\cite{CHR:1997}.  Note also that a gradient of the magnetic
field only leads to a magnetization (but not transport) current.

The noncommutative model above gives a complete expression for the
energy current in terms of the function $c$ appearing in the Poisson
algebra and the energy magnetization $M_E$.  Namely, if we write
$\varepsilon^i=\epsilon^{ij}(\kappa_H E_j+\kappa_H^\mu
\d_j\mu+\kappa_H^T\d_jT+\kappa_H^B\d_jB)$, then
\begin{align}
  \kappa_H &= \frac{\epsilon+p}B\\
  \kappa_H^\mu &= - \frac{Ts+\mu n}B + 
     \Bigl(\frac{\d M_E}{\d \mu}\Bigr)_{T,B},\\
  \kappa_H^T &= - \frac{\mu s}{B}  - cT + 
     \Bigl( \frac{\d M_E}{\d T}\Bigr)_{\mu,B},\\
  \kappa_H^B &=   \Bigl( \frac{\d M_E}{\d B}\Bigr)_{T,\mu}.
\end{align}

\section{Generalized thermal transport}

The Poisson bracket formalism, while
giving a self-consistent set of equations, rely on certain
unjustified assumptions.  For example, the effect of dissipative heat
conduction cannot be taken into account in this formalism.
Fortunately, we can show that beside this effect, the most general
hydrodynamic equations have the same forms as the equations derived
above; the only modification is that there is now no restriction on
the form of $c$, which to this point has been required to be of the
form~(\ref{Jacobi}).

To write down the most general system of hydrodynamic equations, first
we notice that the particle number current cannot be modified due to
the force balance condition.  Thus the only place where modifications
can be made is in the constitutive relation for the energy
current~(\ref{eps-i}).  The dissipative part has the familiar form of
longitudinal heat conduction and shall not be discussed here.  The
most general additional transverse terms one can add to the energy
current is $\epsilon^{ij} \Sigma_a \d_j X^a$
where
$X^a$, $a=1,2,3$, are three independent thermodynamic variables (which 
can be
chosen to be, e.g., $\mu$, $T$, and $B$, but any other choice is
equally valid) and $\Sigma_a=\Sigma_a(X)$ are the corresponding three 
kinetic
coefficients. 
It is convenient to
introduce the one-form $\Sigma\equiv\Sigma_a dX^a$ in the space of
thermodynamic variables.
The constraint on $\Sigma_a$ is
that one can modify the entropy current by adding to Eq.~(\ref{si})
a contribution of the form $\epsilon^{ij}\zeta_a\d_j X^a$
and still preserve the entropy production rate, which should receive 
no contributions
from these new kinetic terms.
By direct calculation using the
thermodynamic relation $ds=T^{-1}(d\varepsilon-\mu dn)$ and the 
conservation
of energy and particle number, 
one can find the divergence of the new entropy current 
\begin{equation}
  \d_t s + \d_i s^i = \frac12 \epsilon^{ij}
   \Bigl(d\zeta - \frac1 Td\Sigma\Bigr)_{ab} \d_i X^a\d_j X^b.
\end{equation}
Therefore, for entropy conservation we need to have
$d\zeta=T^{-1}d\Sigma$.
The most general solution to this equation is $\zeta = d b_0 + c_0dT$,
$\Sigma = d\sigma_0 + Tc_0dT$, where $b_0$, $c_0$ and $\sigma_0$ are
scalar functions (zero-forms) of thermodynamic variables.  In the
energy current, $\sigma_0$ can be absorbed into the magnetization
current, and $c_0$ into $c$ to make the latter a unconstrained
function of three thermodynamic variables. The full energy current
therefore is
\begin{multline}\label{ei}
  \varepsilon^i = \epsilon^{ij} \Bigl[ \frac{\varepsilon+p}B(E_j-
\d_j\mu) -M\d_j \mu +\\
    + \d_j M_E - \cRL T\d_j T \Bigr],
\end{multline}
where $\cRL=c+\mu s/BT$, corresponding to the Righi-Leduc effect with
thermal Hall conductivity $K_H=T\cRL$.  This means that in a
gapped quantum Hall phase at low temperature
$\cRL=\frac\pi6(c_R-c_L)$, where $c_R$ and $c_L$ are the numbers of
right and left moving modes, respectively~\cite{Kane:1997fda,Read:1999fn,Cappelli:2001mp}.

We emphasize here that the fact that the energy current is
parametrized in terms of two functions $c$ and $M_E$ implies one
relationship between the coefficient $\kappa_H^\mu$, $\kappa_H^T$, and
$\kappa_H^B$.  The response to the Luttinger potential coupled to the
energy density~\cite{Luttinger:1964zz} can also be expressed in terms
of these two functions~\cite{GS}.

Despite the fact that the hydrodynamic equations with generic $c$ are
dissipationless (without heat conduction), we are unable to find a
Hamiltonian and a set of Poisson brackets that would lead to these
equations.  We leave the study of the Hamiltonian structure of our
equations to future work.

\subsection{Righi-Leduc coefficient at high temperature}

At low temperature as $\mu$ changes $\cRL$ is expected to vary in a
complicated fashion as the system scans through many quantum Hall
plateaux.  When the temperature is large compared to the interaction
energy, the system is weakly interacting and the Righi-Leduc
coefficient $\cRL$ can be computed reliably.  One can follow the
method of Ref.~\cite{Zyryanov:1964}, but one can also employ the
following short-cut.  In thermal equilibrium, all states in the lowest
Landau level has the same
occupation number $\nu=(e^{-\beta\mu}+1)^{-1}$, which
depends only on $\mu/T$ but not $\mu$ and $T$ separately.  The current
and energy current out of equilibrium, where $\mu$ and $T$ vary in
space, thus depend only on $\mu/T$.  But these quantities have nonzero
dimension, and hence they have to vanish at high temperature.  Thus at
high temperatures and zero electric field, the terms on the right hand
side of Eq.~(\ref{ei}) cancel each other.

The grand partition function in
this regime is
\begin{equation}
  P = \frac{BT}{2\pi}\ln(1+e^{\mu/T}),
\end{equation}
from which all other thermodynamic potentials can be computed.  In
particular, $\varepsilon=p=0$, $M=P/B$.  The condition of vanishing
energy current reads
  $M\d_j\mu  = \d_j M_E -  \cRL T\d_j T$,
which means
\begin{equation}
  \frac{\d M_E}{\d\mu} = M, \qquad
  \frac{\d M_E}{\d T} - T \cRL = 0 .
\end{equation}
The solution to these equations is
\begin{align}
  M_E &= -\frac1{2\pi}T^2 \mathrm{Li}_2 (-e^{\mu/T}) \label{ME-hiT}, 
\\
   \cRL &= -\frac1{2\pi}\left[ \frac\mu T\ln(1+e^{\mu/T}) 
      +2\mathrm{Li}_2(-e^{\mu/T})\right]
  \nonumber\\ & 
= -\frac1{2\pi}\left[\ln\frac\nu{1-\nu}\ln\frac1{1-\nu}
      +2\mathrm{Li}_2\Bigl(-\frac\nu{1-\nu} \Bigr)\right]. \label{cRL-hiT}
\end{align}
The Righi-Leduc coefficient approaches $0$ as $\nu\to0$, and $\pi/6$
as $\nu\to1$.  The latter value matches exactly with that expected for
the $\nu=1$ integer quantum Hall state with a single chiral edge
mode~\cite{Kane:1997fda}.

In the limit of low filling fraction $\nu\ll1$ the formulas simplify.
For example, $\cRL\approx-\frac1{2\pi}(2-\mu/T)e^{\mu/T}$.  It is
interesting to note that, to leading order in $\ln(1/\nu)$,
\begin{equation}
  c= \cRL - \frac{\mu s}{BT} 
  \approx \frac{e^{\mu/T}}{2\pi}\frac{\mu^2}{T^2}
  = \frac{s^2}{nB}\,,
\end{equation}
which is of the form~(\ref{Jacobi}) and moreover coincides with our
initial ``mean-field'' guess for $c$.

\subsection{Particle-hole symmetry}

When the interaction between fermions
is two-body, the system has particle-hole symmetry (for simplicity,
here we assume the magnetic field is uniform and constant).  This
should also be a symmetry of the hydrodynamic equations. 
If one
normalizes the one-body potential so that $\mu=0$ corresponds to a
half-filled Landau level ($\nu=1/2$), then particle-hole symmetry is
the symmetry under $\mu\to-\mu$.  
It is easy to check that the
hydrodynamic equations are particle-hole symmetric if $P$, $M_E$ and
$\cRL$ satisfy
\begin{align}
  P(T,\mu) &= P(T,\mu) - \frac{B\mu}{2\pi}\,,\\
  M_E(T,-\mu) &= \frac1{4\pi}\Bigl(\mu^2+\frac{\mu^2}3T^2\Bigr)
    - M_E(T,\mu) ,\\
  \cRL(T,-\mu) & = \frac\pi6 - \cRL(T,\mu).
\end{align}
In particular, at half filling $\cRL=\pi/12$ if
particle-hole symmetry is not spontaneously broken, which should be
the case at least at sufficiently high temperature. Note that these 
properties
are satisfied by Eqs.~(\ref{ME-hiT}) and (\ref{cRL-hiT}).

\section{Conclusion}

We have shown that the full finite-temperature
hydrodynamics of a system of particles confined to the lowest Landau
level can be written down based on general principles.
We have assumed that the interaction between the electrons are
short-ranged.  In the case of long-ranged Coulomb interaction, we have
to add a Poisson equation for the scalar potential, which can be done
in a straightforward manner.

The Righi-Leduc coefficient $\cRL$ should be viewed as a fundamental
property of a system on the LLL at any filling fraction and
temperature.  In principle, the coefficient can be measured; but the
task may be complicated by the edge transport, as well as from the
contributions from the other terms in the energy current~(\ref{ei}).
We defer a more detailed study to future work.

To the order that we are working on, we are not sensitive to the
first-order corrections to the stress tensor, including the
dissipative shear and bulk viscosities and the dissipationless Hall
viscosity.  These can be introduced
and it would be interesting to investigate their behaviors under
particle-hole symmetry.

\acknowledgments

We thank Sean Hartnoll and Paul Wiegmann 
for discussions.
This work is supported, in part, by the US DOE grant
No.\ DE-FG02-13ER41958, and the ARO-MURI 63834-PH-MUR grant, and a
Simon Investigator grant from the Simons Foundation.
One of the authors (D.~T.~S.) would like to express deep gratitude to
V.~A.~Rubakov for the guidance, inspiration, and
insistence on rigor during this author's time as a graduate student.


\begin{thebibliography}{99}

\bibitem{Tsui:1982yy} 
  D.~C.~Tsui, H.~L.~Stormer, and A.~C.~Gossard,
  Phys.\ Rev.\ Lett.\  {\bf 48}, 1559 (1982).

\bibitem{Laughlin:1983}
  R.~B.~Laughlin,
  Phys.\ Rev.\ Lett.\ {\bf 50}, 1395 (1983).

\bibitem{Cooper:2008}
N.~R.~Copper,
Adv. Phys. {\bf 57}, 539 (2008).

\bibitem{Cooper:2013}
N. R. Cooper and J. Dalibard, 
Phys. Rev. Lett. {\bf 110}, 185301 (2013).

\bibitem{Yao:2013}
N.~Y.~Yao, A.~V.~Gorshkov, C.~R.~Laumann, A.~M.~Läuchli, J.~Ye, and M.~D.~Lukin, 
Phys. Rev. Lett. {\bf 110}, 185302 (2013).

\bibitem{Zyryanov:1964}
P.~S.~Zyryanov,
Phys.\ Stat.\ Sol.\ {\bf 6}, 401 (1964).

\bibitem{Obraztsov:1965}
Yu.~N.~Obraztsov, Sov.\ Phys.\ Solid State {\bf 7}, 455 (1965).

\bibitem{SmrckaStreda:1977}
L.~Smr\v cka and P.~St\v reda,
J.\ Phys.\ C {\bf 10}, 2153 (1977).

\bibitem{CHR:1997}
N.~R.~Cooper, B.~I.~Halperin, and I.~M.~Ruzin,
Phys.\ Rev.\ B {\bf 55}, 2344 (1997).


\bibitem{Girvin:1986zz} 
  S.~M.~Girvin, A.~H.~MacDonald, and P.~M.~Platzman,
  Phys.\ Rev.\ B {\bf 33}, 2481 (1986).

\bibitem{Susskind:2001fb} L.~Susskind,
hep-th/0101029.

\bibitem{Fradkin:2002qw} 
  E.~Fradkin, V.~Jejjala, and R.~G.~Leigh,
  Nucl.\ Phys.\ B {\bf 642}, 483 (2002).

\bibitem{Rubakov_noncomm}
  V.~A.~Rubakov, {\em Classical Field Theories. Theories with Fermions. Noncommutative Theories}, Moscow, URSS (2005) (in Russian).

\bibitem{Landau:1941}
L.~D.~Landau,
J.\ Phys. USSR {\bf 5}, 71 (1941).

\bibitem{Kane:1997fda} 
  C.~L.~Kane and M.~P.~A.~Fisher,
  Phys.\ Rev.\ B {\bf 55}, 15832 (1997).

\bibitem{Stone:2012ud} 
  M.~Stone,
  Phys.\ Rev.\ B {\bf 85}, 184503 (2012).

\bibitem{Read:1999fn} 
  N.~Read and D.~Green,
  Phys.\ Rev.\ B {\bf 61}, 10267 (2000).

\bibitem{Cappelli:2001mp} 
  A.~Cappelli, M.~Huerta and G.~R.~Zemba,
  Nucl.\ Phys.\ B {\bf 636}, 568 (2002).

\bibitem{Simon:1996}
S.~H.~Simon, A.~Stern, and B.~I.~Halperin,
Phys. Rev. B {\bf 54}, 11114(R) (1996).

\bibitem{Streda:1982}
P.~St\v reda,
J.\ Phys.\ C {\bf 15}, L717 (1982).

\bibitem{GSWW}
M.~Geracie, D.~T.~Son, S.-F.~Wu, and C.~Wu,
arXiv:1407.1252.

\bibitem{Luttinger:1964zz} 
  J.~M.~Luttinger,
  Phys.\ Rev.\  {\bf 135}, A1505 (1964).

\bibitem{GS}
  M.~Geracie and D.~T.~Son, arXiv:1408.6843.



\end{thebibliography}
\end{document}